\documentclass[aps,pra,reprint,floatfix,superscriptaddress,nofootinbib]{revtex4-2}
\usepackage{amsmath,amssymb,bm,graphicx,hyperref,microtype}
\usepackage{placeins}
\usepackage{capt-of}
\makeatletter
\newenvironment{fullwidthfigure}{%
  \par\onecolumngrid
  \vspace{6pt}%
}{%
  \par\vspace{6pt}%
  \twocolumngrid\global\@ignoretrue
  \@endpetrue
}
\makeatother
\graphicspath{{figures/}}
\hypersetup{colorlinks=true,linkcolor=blue,citecolor=blue,urlcolor=blue}

\newcommand{\dd}{\mathrm d}

\newcommand{\nb}{\bar n}

\begin{document}

\title{Memory and thermal amplification in spin--cavity squared commutators}

\author{Yong-Hong Ma}
\affiliation{School of Science, Inner Mongolia University of Science and Technology, Baotou 014010, China}
\author{Hui-Hui Xu}
\affiliation{School of Science, Inner Mongolia University of Science and Technology, Baotou 014010, China}
\author{Jian-Zhuang Wu}
\affiliation{Center for Quantum Sciences and School of Physics, Northeast Normal University, Changchun 130117, China}
\author{Quan-Zhen Ding}
\affiliation{Center for Controlled Quantum Systems and Department of Physics and Engineering Physics, Stevens Institute of Technology, Hoboken, New Jersey 07030, USA}
\author{Wu-Ming Liu}
\affiliation{School of Science, Inner Mongolia University of Science and Technology, Baotou 014010, China}
\author{Xin-Yu Zhao}
\email{xzhao1@foxmail.com}
\affiliation{Department of Physics, Fuzhou University, Fuzhou 350116, People’s Republic of China}
\author{E Wu}
\email{towue@163.com}
\affiliation{School of Science, Inner Mongolia University of Science and Technology, Baotou 014010, China}

\begin{abstract}
Squared commutators in the Holstein--Primakoff limit of a spin--cavity system
provide a compact way to separate propagation from covariance growth in a
finite-temperature reservoir with memory.  In the finite-temperature NMQSD construction, the linear quadrature commutator is fixed by the retarded
spin--cavity propagator, whereas a quadratic commutator carries the same
retarded factor together with a covariance factor.  For a zero-mean Gaussian
state,
\(C_{R_i^2,R_j}(t)=4|\kappa_{ij}(t)|^2V_{ii}(t)\); the symmetrized expression gives
the spin-side and mixed channels.  Since \(\bar n\) enters the covariance sector
but not the homogeneous retarded kernel, raising \(\bar n\) from 0 to 1 leaves
the linear transfer unchanged while increasing the quadratic signal.  Varying
the bath-memory rate and the counter-rotating coupling within the stable HP
region then shows how stored cavity history changes both the transfer weight and
its distribution in time.  The calculation separates memory-dependent
propagation from thermal covariance growth in collective spin--cavity dynamics.
\end{abstract}

\maketitle

\section{Introduction}

Squared commutators and out-of-time-order correlators are commonly used to study operator growth and scrambling in closed many-body systems~\cite{LarkinOvchinnikov1969,MaldacenaShenkerStanford2016,HashimotoMurataYoshii2017,Swingle2018,XuSwingle2024,Xu:25,PhysRevE.98.012216,PhysRevA.101.053604,PhysRevB.98.134305,PhysRevB.101.174313,PhysRevE.98.012216}.  In an open system, the object must be specified more carefully.  A reduced-system commutator, a measurement protocol, and a full system--environment correlator are different quantities~\cite{SyzranovGorshkovGalitski2018,Tuziemski2019,6fsj-6nc5,98vl-zlcs,PhysRevLett.112.246401,PhysRevB.98.184416,PhysRevB.88.075207}.  The reduced quadrature commutator considered here is the retarded response between two canonical variables of a Gaussian spin--cavity model~\cite{Kubo1957,MollabashiRahimiKeshari2025,Weedbrook2012,PhysRevB.103.174106,x19r-pzyb,PhysRevB.100.014415,PhysRevResearch.7.L012058,PhysRevA.81.062308,PhysRevB.104.L100415}.  The question is how a reservoir with a finite correlation time changes the transfer of an operator between the cavity and the collective spin.

The Dicke model is a natural collective spin--cavity setting for this problem~\cite{Dicke1954,TavisCummings1968,DimerEstienneParkinsCarmichael2007,BaumannGuerlinBrenneckeEsslinger2010,RitschDomokosBrenneckeEsslinger2013,KirtonKeelingBrandes2019,ChavezCarlos2019,TiwariBanerjee2023,PhysRevD.109.L121902}.  In optical-cavity realizations, the external reservoir usually acts on the cavity field.  In the Holstein--Primakoff (HP) regime~\cite{HolsteinPrimakoff1940,EmaryBrandes2003,PhysRevA.81.032314,gckn-mc7w,PhysRevResearch.2.043243,PhysRevA.39.3196,PhysRevA.74.054101,PhysRevA.86.012116}, a polarized collective spin is represented by a bosonic fluctuation mode $b$.  Keeping both the exchange and counter-rotating couplings to the cavity gives a quadratic Hamiltonian, so the finite-temperature non-Markovian quantum-state-diffusion (NMQSD) $O_1/O_2$ equations and the retarded commutator can be obtained explicitly.

A perturbation placed in one cavity quadrature at $t=0$ can be followed through its response in a spin quadrature~\cite{PhysRevA.96.032125,PhysRevA.84.043819,PhysRevA.96.063820,PhysRevResearch.4.023101,PhysRevA.104.013722,PhysRevB.111.104425,PhysRevB.107.045425}.  When the bath memory is short, the cavity damping is nearly local; as the correlation time increases, the cavity equation retains an earlier part of its own motion.  This stored history changes both the amplitude and timing of the cross commutator.  The counter-rotating coupling $G_r$ is also varied, since it shifts the two quadrature sectors in opposite directions.

The Gaussian model lets us keep propagation and state fluctuations apart.  The bath commutator fixes the retarded propagator, whereas the symmetrized bath noise changes the covariance matrix; in a quadratic squared commutator the two contributions appear in the same signal but through different factors.  When the reservoir has both finite memory and finite temperature, that separation is useful because propagation and covariance growth are otherwise easy to mix.

For linear quadratures, the squared commutator is the square of a retarded transfer coefficient, while replacing one quadrature by its square keeps this coefficient and adds a covariance.  Temperature can change the quadratic signal even when the linear transfer is unchanged.  The derivation uses finite-temperature NMQSD~\cite{DiosiGisinStrunz1998,StrunzDiosiGisin1999,YuDiosiGisinStrunz1999,Yu2004,ChenYu2014,PhysRevA.87.052127,PhysRevA.90.042108,PhysRevA.88.052122}; related finite-temperature and non-Markovian QSD methods have also been used for spin squeezing~\cite{MaDingYu2020,DingZhaoMaChen2021,ChenDingShiJunYu2020}.  Scans over the bath-memory rate and $G_r$ then show how the retarded part of the signal is redistributed inside the stable HP region.

\section{Model}

The HP spin--cavity Hamiltonian used here is
\begin{equation}
H_{\rm HP}=\omega_c c^\dagger c+\omega_s b^\dagger b
+G(cb^\dagger+c^\dagger b)+G_r(c^\dagger b^\dagger+cb).
\label{eq:HHP}
\end{equation}
Here $c$ is the cavity mode and $b$ is the HP spin-fluctuation mode.  The coupling $G$ exchanges excitations, while $G_r$ is the counter-rotating pairing term.  We treat them as independent effective parameters.  The choice $G_r=0$ gives the exchange limit; finite $G_r$ mixes creation and annihilation processes.

Using
\begin{align}
X_c&=\frac{c+c^\dagger}{\sqrt2},&
P_c&=\frac{c-c^\dagger}{i\sqrt2},\nonumber\\
X_s&=\frac{b+b^\dagger}{\sqrt2},&
P_s&=\frac{b-b^\dagger}{i\sqrt2},
\label{eq:quadratures}
\end{align}
Eq.~\eqref{eq:HHP} separates into two quadratic sectors,
\begin{align}
H_X&=\frac{\omega_c}{2}X_c^2+\frac{\omega_s}{2}X_s^2+(G+G_r)X_cX_s,\nonumber\\
H_P&=\frac{\omega_c}{2}P_c^2+\frac{\omega_s}{2}P_s^2+(G-G_r)P_cP_s,
\label{eq:quadrature_sectors}
\end{align}
up to constants.  The stable Gaussian region is defined by
\begin{align}
\Delta_X&=\omega_c\omega_s-(G+G_r)^2>0,\nonumber\\
\Delta_P&=\omega_c\omega_s-(G-G_r)^2>0.
\label{eq:stable}
\end{align}
The stable Gaussian region is bounded by these two inequalities.  As one margin becomes small, the corresponding quadrature sector softens and the transfer can grow; all main scans stay inside this region, with the maps showing the distance to its boundary.

The covariance also gives a direct check of the HP approximation.  With the normalization in Eq.~\eqref{eq:quadratures}, the spin-wave occupation is
\begin{equation}
n_s(t)=\langle b^\dagger b\rangle=\frac{V_{X_sX_s}(t)+V_{P_sP_s}(t)-1}{2}.
\label{eq:ns_HP_control}
\end{equation}
At fixed collective couplings, the HP expansion is controlled by \(n_s^{\rm max}/N\).  Values for the parameter sets used in the figures are listed in the Supplemental Material.

Equation~\eqref{eq:quadrature_sectors} also makes the role of $G_r$ explicit.  It enters the $X$ and $P$ sectors with opposite signs, so the two sector frequencies move in opposite directions.  Since a single peak height does not describe this redistribution well, we also use the integrated transfer and its time centroid.

Figure~\ref{fig:model} summarizes the model and the calculation.  The spin and cavity form two coupled quadrature sectors: $G$ enters both with the same sign, whereas $G_r$ enters with opposite signs.  Because the reservoir acts only on the cavity, its memory reaches the spin through the cavity field.  A weak local spin decay would add drift and diffusion, but it would not change the factorized structure derived below.  The cavity-to-spin transfer signal is $C_{P_sX_c}$.

\begin{fullwidthfigure}
\centering
\includegraphics[width=0.86\textwidth]{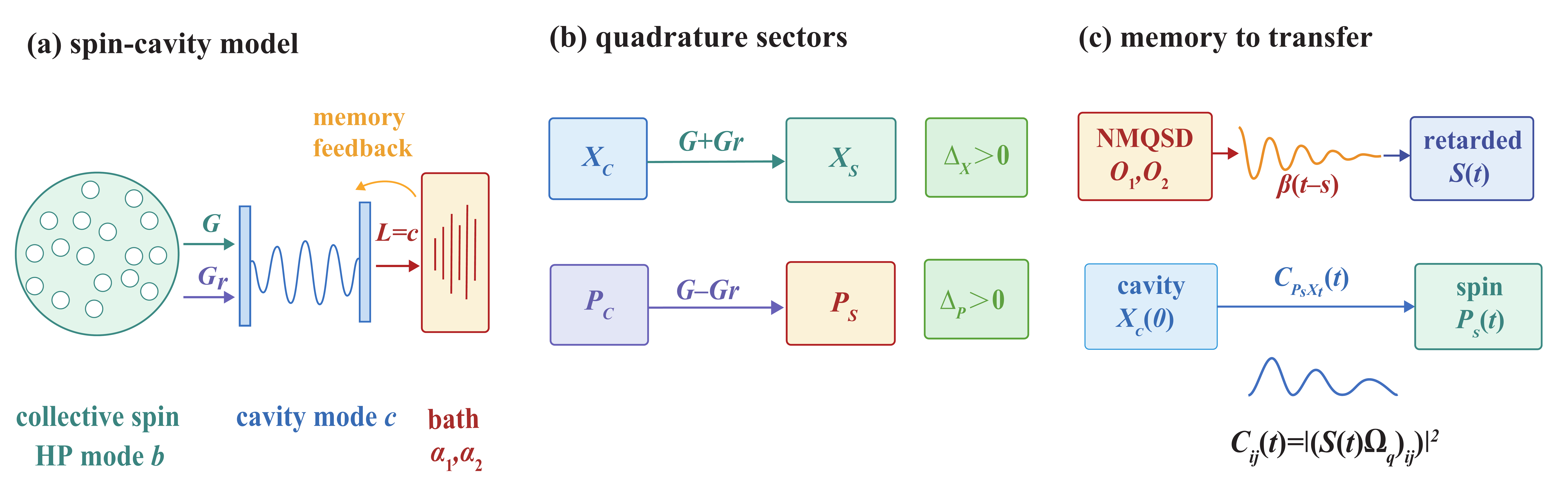}
\captionof{figure}{Model and calculation route.  (a) The HP spin mode $b$ couples to the cavity mode $c$, and the reservoir acts through $L=c$.  (b) The $X$ and $P$ sectors contain the couplings $G+G_r$ and $G-G_r$ and have stability margins $\Delta_X$ and $\Delta_P$.  (c) The two finite-temperature NMQSD correlations give the retarded kernel $\beta$ and the propagator $S(t)$.  The linear channel contains the retarded factor alone.  The quadratic channel also contains a covariance.\label{fig:model}}

\end{fullwidthfigure}

\section{Finite-temperature NMQSD construction}

The reservoir is coupled through
\begin{equation}
H_{\rm int}=LB^\dagger+L^\dagger B,\qquad L=c.
\end{equation}
Finite-temperature NMQSD requires two bath correlations,
\begin{align}
\alpha_1(t,s)&=\langle B(t)B^\dagger(s)\rangle,
&
\alpha_2(t,s)&=\langle B^\dagger(t)B(s)\rangle,
\end{align}
which contain the emission and absorption factors \(\bar n(\omega)+1\) and \(\bar n(\omega)\).  The linear QSD equation is
\begin{align}
\partial_t\psi_t={}&
\left[-iH_{\rm HP}+cz_1^*(t)+c^\dagger z_2^*(t)\right.\nonumber\\
&\left.-c^\dagger\bar O_1(t)-c\bar O_2(t)\right]\psi_t.
\label{eq:qsd}
\end{align}
The boundary values are \(O_1(t,t)=c\) and \(O_2(t,t)=c^\dagger\), with
\begin{equation}
O_\mu(t,s)\psi_t=\frac{\delta\psi_t}{\delta z_\mu^*(s)},\qquad
\bar O_\mu(t)=\int_0^t\dd s\,\alpha_\mu(t,s)O_\mu(t,s).
\end{equation}

For this quadratic model, each functional-derivative operator contains a linear part and an identity part.  With \(\bm a=(c,c^\dagger,b,b^\dagger)^T\),
\begin{equation}
\begin{aligned}
O_\mu(t,s)&=\bm f_\mu^T(t,s)\bm a\\
&\quad+\sum_{\nu=1}^2\int_s^t\dd u\,z_\nu^*(u)h_{\mu\nu}(t,s;u)I .
\end{aligned}
\label{eq:O_ansatz}
\end{equation}
The linear coefficients determine the retarded kernel, while the identity term carries the finite-temperature noise history required by the NMQSD consistency equation.  The resulting coefficient equations are given in the Supplemental Material.

For the analytic colored-reservoir kernel used below,
\begin{align}
\alpha_1(t,s)&=\frac{\Gamma_c\gamma_B}{2}(\nb+1)e^{-\gamma_B(t-s)},\nonumber\\
\alpha_2(t,s)&=\frac{\Gamma_c\gamma_B}{2}\nb e^{-\gamma_B(t-s)},\qquad t\ge s.
\label{eq:alpha_ou}
\end{align}
The correlation time of this one-pole reservoir is \(\gamma_B^{-1}\).  With this form the memory integral remains simple, and the bath time can be compared directly with the spin--cavity transfer time.  The identity-sector update contains the time-ordered partial memory history
\begin{equation}
\bar{\bm f}^{\le u}_\mu(t)=\int_0^u\dd s\,\alpha_\mu(t,s)\bm f_\mu(t,s).
\label{eq:partialf}
\end{equation}
The terminal time remains $t$, but the history stops at the differentiation time $u$, as required by the time ordering of the functional derivative.  In general, this quantity differs from both $\bar{\bm f}_\mu(t)$ and $\bar{\bm f}_\mu(u)$.

At the quadrature level the same reduced dynamics can be written as
\begin{equation}
\bm r(t)=S(t)\bm r(0)+\int_0^t\dd\tau\,\mathcal G(t,\tau)\bm\xi(\tau),
\label{eq:gaussian_solution}
\end{equation}
where \(\mathcal G(t,\tau)\) propagates cavity noise injected at time \(\tau\) to the system quadratures at time \(t\).  We take a factorized system--reservoir state at $t=0$ and read the following commutator as the response to a perturbation applied at that time.  Then
\begin{equation}
[r_i(t),r_j(0)]=i[S(t)\Omega_q]_{ij},
\label{eq:linear_sees_retarded}
\end{equation}
so the linear commutator contains only the homogeneous propagator.  The covariance also contains the symmetrized cavity-noise kernel
\begin{align}
\frac{1}{2}\langle \bm\xi(t)\bm\xi^T(s)+\bm\xi(s)\bm\xi^T(t)\rangle
&=D_\xi(t-s)\Pi_c,\nonumber\\
\Pi_c&=\mathrm{diag}(1,1,0,0),\nonumber\\
D_\xi(\tau)&=\frac{\Gamma_c\gamma_B}{4}(2\bar n+1)e^{-\gamma_B|\tau|},
\label{eq:Dxi_main}
\end{align}
where \(\Pi_c\) projects onto \((X_c,P_c)\).  Initial system--reservoir correlations can change the early covariance transient.  They do not change the definition of the homogeneous transfer amplitude in Eq.~\eqref{eq:linear_sees_retarded}.  The covariance is
\begin{align}
V(t)=&\,S(t)V_0S^T(t)\nonumber\\
&+\int_0^t\dd\tau\int_0^t\dd\tau'\,
\mathcal G(t,\tau)D_\xi(\tau-\tau')\Pi_c\mathcal G^T(t,\tau').
\label{eq:covariance_main}
\end{align}

\section{Linear and quadratic squared commutators}

\subsection{Linear channel}

For the linear quadrature commutator, the homogeneous retarded kernel is
\begin{equation}
\beta(t,s)=\alpha_1(t,s)-\alpha_2(s,t).
\label{eq:beta}
\end{equation}
For stationary correlations, the reversed argument in \(\alpha_2(s,t)\) denotes the same bath correlation evaluated on the reverse time ordering.
For Eq.~\eqref{eq:alpha_ou},
\begin{equation}
\beta(t,s)=\frac{\Gamma_c\gamma_B}{2}e^{-\gamma_B(t-s)},\qquad t\ge s.
\label{eq:beta_ou}
\end{equation}
The thermal occupation cancels from the homogeneous commutator propagation.  The result is the finite-memory version of the usual Markovian separation between first-moment damping and thermal diffusion~\cite{GoriniKossakowskiSudarshan1976,Lindblad1976}; the occupation still enters the covariance and therefore affects state-dependent quadratic channels.

Introducing auxiliary variables $y_c,y_c^\dagger$, the retarded operator equations are
\begin{align}
\dot c&=-i\omega_c c-iG b-iG_r b^\dagger-y_c,\nonumber\\
\dot c^\dagger&=+i\omega_c c^\dagger+iG b^\dagger+iG_r b-y_c^\dagger,\nonumber\\
\dot b&=-i\omega_s b-iG c-iG_r c^\dagger,\nonumber\\
\dot b^\dagger&=+i\omega_s b^\dagger+iG c^\dagger+iG_r c,\nonumber\\
\dot y_c&=\frac{\Gamma_c\gamma_B}{2}c-\gamma_B y_c,
\qquad
\dot y_c^\dagger=\frac{\Gamma_c\gamma_B}{2}c^\dagger-\gamma_B y_c^\dagger.
\label{eq:retarded_eqs}
\end{align}
The two auxiliary variables rewrite the memory convolution as local equations.  For example,
\begin{equation}
y_c(t)=\int_0^t\dd s\,\frac{\Gamma_c\gamma_B}{2}e^{-\gamma_B(t-s)}c(s),
\label{eq:yc_convolution}
\end{equation}
and similarly for \(c^\dagger\).  In the short-memory limit, \(y_c(t)=\Gamma_c c(t)/2+O(\gamma_B^{-1})\), and the damping becomes local.  For finite \(\gamma_B\), the same variable stores a filtered part of the earlier cavity motion.  The relevant comparison is between the spin--cavity period and \(\gamma_B^{-1}\).  We use \(\gamma_B=50\) as the short-memory reference and \(\gamma_B=0.5\) as a case in which the bath correlation time is comparable to the transfer time.  In this paper, ``memory'' refers to this finite correlation time.

The physical spin--cavity sector remains four dimensional; the two extra variables only store the reservoir history.  We obtain the retarded curves by exponentiating the constant $6\times6$ matrix and taking its physical $4\times4$ block.

If $\bm r=(X_c,P_c,X_s,P_s)^T$ and $\bm r(t)=S(t)\bm r(0)$, then
\begin{equation}
[r_i(t),r_j(0)]=i[S(t)\Omega_q]_{ij},
\end{equation}
where
\begin{equation}
\Omega_q=\begin{pmatrix}
0&1&0&0\\
-1&0&0&0\\
0&0&0&1\\
0&0&-1&0
\end{pmatrix}.
\end{equation}
The squared commutator is therefore
\begin{equation}
C_{ij}(t)=\left|[S(t)\Omega_q]_{ij}\right|^2.
\label{eq:Cij}
\end{equation}
Equivalently, with
\begin{equation}
\kappa_{ij}(t)=[S(t)\Omega_q]_{ij},
\label{eq:kappa}
\end{equation}
the linear Gaussian squared commutator is
\begin{equation}
C_{R_iR_j}(t)=|\kappa_{ij}(t)|^2.
\label{eq:linear_kappa}
\end{equation}
We use both $C_{P_sX_c}$ and $C_{X_cP_s}$: the first follows transfer from the cavity to the spin, whereas the second pairs naturally with the quadratic operator $X_c^2$.  Their magnitudes are equal for the symmetric parameters used in the representative traces.  Since the linear commutator is a c-number in this Gaussian model, its value does not depend on the initial covariance.

\begin{fullwidthfigure}
\centering
\includegraphics[width=0.86\textwidth]{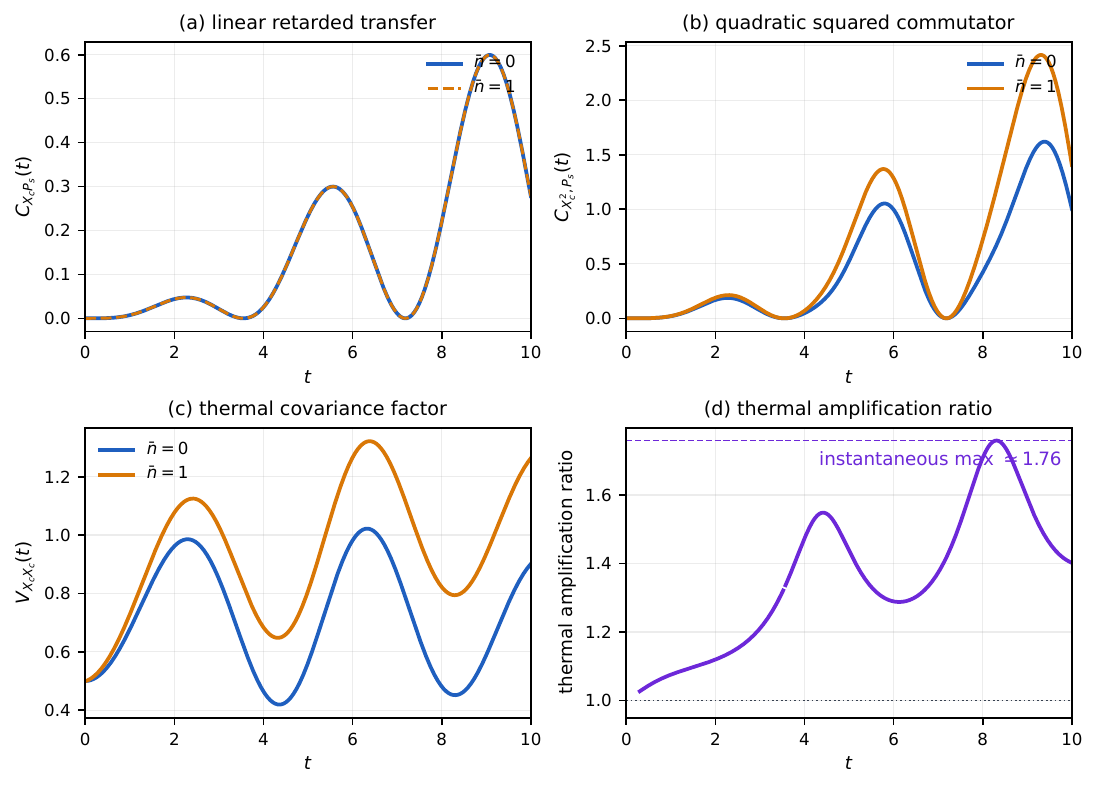}
\captionof{figure}{Linear and quadratic channels at two bath occupations.  Parameters are \(\omega_c=\omega_s=1\), \(V_0=I_4/2\), \(G=0.1\), \(G_r=0.5\), \(\Gamma_c=0.05\), and \(\gamma_B=50\).  (a) The linear transfer \(C_{X_cP_s}\) is the same for \(\bar n=0\) and 1.  (b) The quadratic channel \(C_{X_c^2,P_s}\) increases.  (c) The change comes from \(V_{X_cX_c}\).  (d) Ratio of the two quadratic curves, excluding points near zeros of the denominator.  Its largest value is about 1.76, and the ratio of peak amplitudes is about 1.49.  For this trace, \(n_s^{\rm max}\simeq0.44\) at \(\bar n=1\).\label{fig:quadratic}}

\end{fullwidthfigure}

\subsection{Quadratic channel}

To make the covariance contribution visible, we use a quadratic operator in
one slot of the squared commutator.  Since
\begin{equation}
[R_i(t),R_j(0)]=i\kappa_{ij}(t)
\end{equation}
is a c-number in the Gaussian theory,
\begin{equation}
\begin{aligned}
{}[R_i^2(t),R_j(0)]
&=R_i(t)[R_i(t),R_j(0)]\\
&\quad +[R_i(t),R_j(0)]R_i(t)\\
&=2i\kappa_{ij}(t)R_i(t).
\end{aligned}
\label{eq:quadratic_commutator}
\end{equation}
For a zero-mean Gaussian state, define the time-dependent covariance
\begin{equation}
V_{ii}(t)\equiv \langle R_i^2(t)\rangle .
\label{eq:Vii_def}
\end{equation}
For off-diagonal entries we use \(V_{ik}(t)=\langle\{R_i(t),R_k(t)\}\rangle/2\).
The same quantity is the Schr\"odinger-picture covariance of the evolved Gaussian state.  In the Heisenberg picture, the average is taken over the initial state and the bath noise; the state is not evolved a second time.  The corresponding squared commutator is
\begin{equation}
\begin{aligned}
C_{R_i^2,R_j}(t)
&=\left\langle [R_i^2(t),R_j(0)]^\dagger
[R_i^2(t),R_j(0)]\right\rangle\\
&=4|\kappa_{ij}(t)|^2V_{ii}(t).
\end{aligned}
\label{eq:quadratic_factorization}
\end{equation}
The quadratic channel contains the same retarded coefficient as the linear channel, but it is multiplied by a state covariance.  Equation~\eqref{eq:quadratic_factorization} separates the measured signal into a propagation part and a covariance part.  For the main spin--cavity example,
\begin{equation}
C_{X_c^2,P_s}(t)=4C_{X_cP_s}(t)V_{X_cX_c}(t).
\label{eq:Xc2Ps_factorization}
\end{equation}
The memory kernel and $G_r$ determine the retarded factor $C_{X_cP_s}$, whereas thermal diffusion changes the covariance $V_{X_cX_c}$.  Their product gives the curve in Fig.~\ref{fig:quadratic}.  The same algebra applies to spin-side and mixed quadratic operators, whose formulas and traces are given in the Supplemental Material.

Unless stated otherwise, the numerical examples use \(\omega_c=\omega_s=1\),
which sets the frequency unit, a zero-mean bare two-mode vacuum covariance
\(V_0=I_4/2\), and the observation window \(0\le t\le T=10\).

For all stable support families, the peak linear transfer is unchanged between $\bar n=0$ and 1 within numerical precision, whereas the quadratic peak changes.  Its ratio is 1.49 in Fig.~\ref{fig:quadratic}.  An independent computation of the linear commutator and the covariance gives the same factorization in Eq.~\eqref{eq:Xc2Ps_factorization}.  In the later scans, memory and $G_r$ change the retarded factor, while the occupation $\bar n$ enters through the covariance.

The spin-side and mixed quadratic channels show the same temperature dependence.  We place these curves in the Supplemental Material and keep the main text focused on Eq.~\eqref{eq:Xc2Ps_factorization} and the memory scans.

To characterize a full transfer pulse rather than a single oscillation lobe, we also use the accumulated transfer and its centroid,
\begin{equation}
A_C=\int_0^T\!\dd t\,C_{P_sX_c}(t),
\qquad
 t_{\rm cm}=\frac{\int_0^T\!\dd t\,t\,C_{P_sX_c}(t)}{\int_0^T\!\dd t\,C_{P_sX_c}(t)} .
\label{eq:area_centroid}
\end{equation}
When the curve has several lobes, the global peak time can jump from one lobe to another.  The area and centroid are more stable measures.  In the memory maps we use the relative area change
\begin{equation}
\Delta A_C=\frac{A_C(\gamma_B=0.5)-A_C(\gamma_B=50)}{A_C(\gamma_B=50)}
\end{equation}
and the centroid shift \(\Delta t_{\rm cm}=t_{\rm cm}(\gamma_B=0.5)-t_{\rm cm}(\gamma_B=50)\).  Peak height is useful for individual traces, while $A_C$ and $t_{\rm cm}$ are more stable for broad parameter maps.

\section{Dependence on bath memory}

\begin{fullwidthfigure}
\centering
\includegraphics[width=0.86\textwidth]{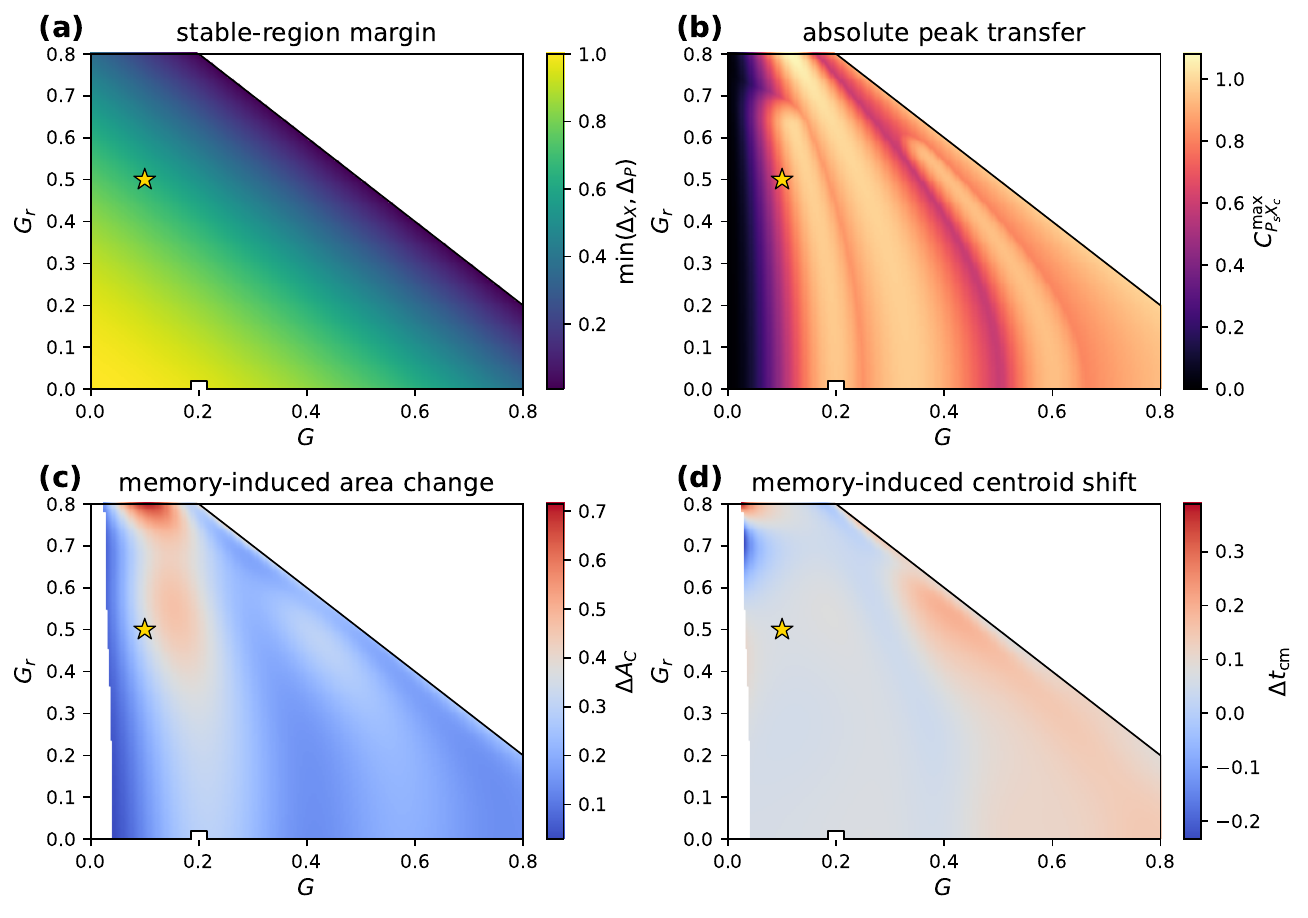}
\captionof{figure}{Transfer inside the stable HP region.  (a) Stability margin $\min(\Delta_X,\Delta_P)$.  The star marks the point used in Fig.~\ref{fig:memory}; the square marks the $G=0.2$ line used in Fig.~\ref{fig:counter}.  (b) Peak $C_{P_sX_c}$ for $\gamma_B=0.5$.  (c) Relative area change between $\gamma_B=0.5$ and 50.  (d) Centroid shift over the same comparison.  The centroid is omitted where the transfer is too weak.  Blank points inside the stable region are weak-transfer points, not missing interpolation data.\label{fig:landscape}}

\end{fullwidthfigure}

Figure~\ref{fig:landscape} gives the stable-region overview.  Panel (b) shows the peak transfer for $\gamma_B=0.5$.  Panels (c) and (d) compare this case with the short-memory reference $\gamma_B=50$ through the area and centroid.  We mask the centroid where the transfer is too weak to define it reliably.  On the remaining points, the area increase ranges from about 7\% to 72\%, with a median near 11\%.  The centroid shift lies between about $-0.23$ and 0.39.  The point used in Fig.~\ref{fig:memory} lies inside this broader response region.

\subsection{Time-domain curves}

\begin{fullwidthfigure}
\centering
\includegraphics[width=0.86\textwidth]{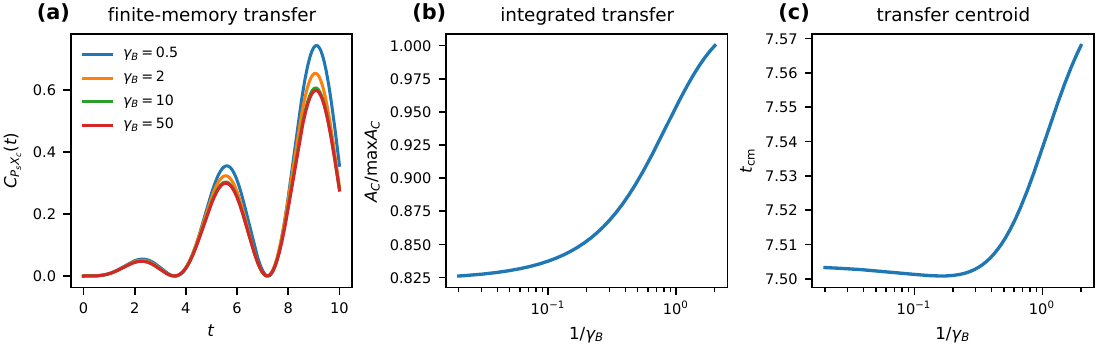}
\captionof{figure}{Time-domain transfer for different bath-memory rates.  The parameters are $\omega_c=\omega_s=1$, $G=0.1$, $G_r=0.5$, and $\Gamma_c=0.05$.  (a) $C_{P_sX_c}(t)$ for several $\gamma_B$ values; smaller $\gamma_B$ means a longer correlation time.  (b) Integrated area, normalized by its largest value in the sweep.  (c) Transfer centroid $t_{\rm cm}$.\label{fig:memory}}

\end{fullwidthfigure}

Figure~\ref{fig:memory} shows that changing $\gamma_B$ affects both the size and the timing of the transfer.  At large $\gamma_B$, the auxiliary variable follows the cavity almost immediately; for smaller $\gamma_B$, it retains part of the earlier cavity motion, so the spin is driven by a field with a finite history.  The area and centroid summarize this change over the full time window.

\section{Effect of the counter-rotating coupling}

\begin{fullwidthfigure}
\centering
\includegraphics[width=0.86\textwidth]{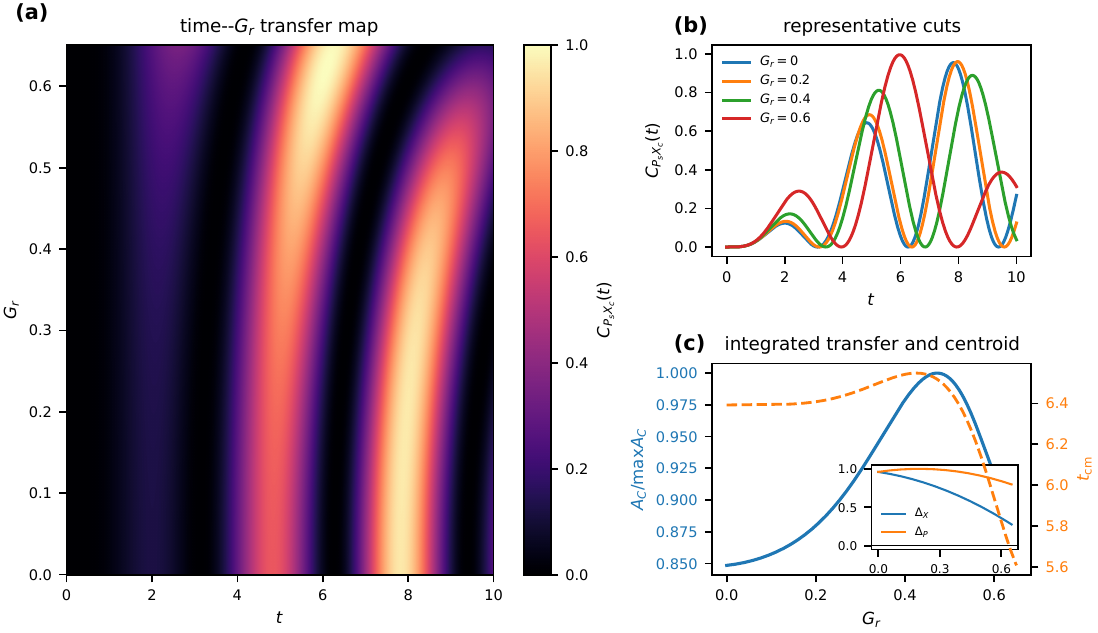}
\captionof{figure}{Dependence on the counter-rotating coupling.  The sweep uses $\omega_c=\omega_s=1$, $G=0.2$, $\Gamma_c=0.05$, and $\gamma_B=0.5$.  The value of $G_r$ is varied inside the stable region.  (a) Time--$G_r$ map of $C_{P_sX_c}(t)$.  (b) Selected time traces.  (c) Integrated area and centroid, with the two stability margins $\Delta_X$ and $\Delta_P$.\label{fig:counter}}

\end{fullwidthfigure}

Changing $G_r$ shifts the two sector frequencies, as seen in Eq.~\eqref{eq:quadrature_sectors}, but the result is not a simple overall gain.  The two sectors contribute with different phases, so the transfer is redistributed across the observation window.  The area and centroid in Fig.~\ref{fig:counter} show this change over the full window.

\section{Discussion and conclusion}

Temperature does not enter every squared commutator in the same way.  The linear spin--cavity commutator is the retarded response of the HP spin mode to a cavity perturbation, with its value fixed by the memory kernel and the two quadrature sectors.  A quadratic operator probes that response after it has been dressed by one or more covariances.  This is the relation behind $C_{X_c^2,P_s}$ in the main text and behind $C_{n_s,X_c}$ and $C_{\{X_c,X_s\}/2,P_s}$ in the Supplemental Material.

The bath occupation changes the covariances through diffusion while dropping out of the homogeneous retarded kernel.  The linear curves are therefore independent of $\bar n$, whereas the quadratic curves vary with $\bar n$.  The memory rate and $G_r$ control the propagation part of the signal; the occupation controls its covariance dressing.

For a quadratic Hamiltonian with linear bath coupling, $[r_i(t),r_j(0)]/i$ follows the same linear flow as the corresponding classical Poisson bracket.  We use this correspondence as a check: the finite-temperature $O_1/O_2$ construction agrees with the retarded Green function and the covariance equation~\cite{GardinerZoller2004,PhysRevD.87.024011,PhysRevE.70.015601,WallsMilburn2008,PhysRevLett.119.176403,PhysRevA.94.062512,PhysRevB.85.035115,BreuerPetruccione2002}.  The coefficient equations, the partial-memory update, and the numerical tests are given in the Supplemental Material.

Beyond the HP approximation, spin saturation and nonlinear operator growth can mix propagation with state fluctuations, so the exact product form may be lost.  The Gaussian result is then a reference for identifying these nonlinear corrections.
\section*{Acknowledgements}
This work was partly supported by the National Natural Science Foundation of China under Grant No. 12265022, the Inner Mongolia Natural Science Foundation under Grant No. 2025MS01005 and No. 2026MS0476, and Elite Revitalizing Inner Mongolia Program (2025TGL05).
\bibliographystyle{apsrev4-2}
\bibliography{references}

@article{Dicke1954,
  author = {Dicke, R. H.},
  title = {Coherence in Spontaneous Radiation Processes},
  journal = {Physical Review},
  volume = {93},
  pages = {99--110},
  year = {1954},
  doi = {10.1103/PhysRev.93.99}
}

@article{TavisCummings1968,
  author = {Tavis, Michael and Cummings, Frederick W.},
  title = {Exact Solution for an {N}-Molecule--Radiation-Field Hamiltonian},
  journal = {Physical Review},
  volume = {170},
  pages = {379--384},
  year = {1968},
  doi = {10.1103/PhysRev.170.379}
}

@article{HolsteinPrimakoff1940,
  author = {Holstein, T. and Primakoff, H.},
  title = {Field Dependence of the Intrinsic Domain Magnetization of a Ferromagnet},
  journal = {Physical Review},
  volume = {58},
  pages = {1098--1113},
  year = {1940},
  doi = {10.1103/PhysRev.58.1098}
}

@article{EmaryBrandes2003,
  author = {Emary, Clive and Brandes, Tobias},
  title = {Chaos and the quantum phase transition in the Dicke model},
  journal = {Physical Review E},
  volume = {67},
  pages = {066203},
  year = {2003},
  doi = {10.1103/PhysRevE.67.066203}
}

@article{DiosiGisinStrunz1998,
  author = {Di{\'o}si, Lajos and Gisin, Nicolas and Strunz, Walter T.},
  title = {Non-Markovian quantum state diffusion},
  journal = {Physical Review A},
  volume = {58},
  pages = {1699--1712},
  year = {1998},
  doi = {10.1103/PhysRevA.58.1699}
}

@article{StrunzDiosiGisin1999,
  author = {Strunz, Walter T. and Di{\'o}si, Lajos and Gisin, Nicolas},
  title = {Open System Dynamics with Non-Markovian Quantum Trajectories},
  journal = {Physical Review Letters},
  volume = {82},
  pages = {1801--1805},
  year = {1999},
  doi = {10.1103/PhysRevLett.82.1801}
}

@article{YuDiosiGisinStrunz1999,
  author = {Yu, Ting and Di{\'o}si, Lajos and Gisin, Nicolas and Strunz, Walter T.},
  title = {Non-Markovian quantum-state diffusion: Perturbation approach},
  journal = {Physical Review A},
  volume = {60},
  pages = {91--103},
  year = {1999},
  doi = {10.1103/PhysRevA.60.91}
}

@article{Yu2004,
  author = {Yu, Ting},
  title = {Non-Markovian quantum trajectories versus master equations: Finite-temperature heat bath},
  journal = {Physical Review A},
  volume = {69},
  pages = {062107},
  year = {2004},
  doi = {10.1103/PhysRevA.69.062107}
}

@article{MaldacenaShenkerStanford2016,
  author = {Maldacena, Juan and Shenker, Stephen H. and Stanford, Douglas},
  title = {A bound on chaos},
  journal = {Journal of High Energy Physics},
  volume = {2016},
  pages = {106},
  year = {2016},
  doi = {10.1007/JHEP08(2016)106}
}

@article{HashimotoMurataYoshii2017,
  author = {Hashimoto, Koji and Murata, Keiju and Yoshii, Ryosuke},
  title = {Out-of-time-order correlators in quantum mechanics},
  journal = {Journal of High Energy Physics},
  volume = {2017},
  pages = {138},
  year = {2017},
  doi = {10.1007/JHEP10(2017)138}
}

@article{Swingle2018,
  author = {Swingle, Brian},
  title = {Unscrambling the physics of out-of-time-order correlators},
  journal = {Nature Physics},
  volume = {14},
  pages = {988--990},
  year = {2018},
  doi = {10.1038/s41567-018-0295-5}
}

@article{XuSwingle2024,
  author = {Xu, Shenglong and Swingle, Brian},
  title = {Scrambling Dynamics and Out-of-Time-Ordered Correlators in Quantum Many-Body Systems},
  journal = {PRX Quantum},
  volume = {5},
  pages = {010201},
  year = {2024},
  doi = {10.1103/PRXQuantum.5.010201}
}

@article{SyzranovGorshkovGalitski2018,
  author = {Syzranov, Sergey V. and Gorshkov, Alexey V. and Galitski, Victor},
  title = {Out-of-time-order correlators in finite open systems},
  journal = {Physical Review B},
  volume = {97},
  pages = {161114},
  year = {2018},
  doi = {10.1103/PhysRevB.97.161114}
}

@article{ChavezCarlos2019,
  author = {Ch{\'a}vez-Carlos, Jorge and L{\'o}pez-del-Carpio, Bernardo and Bastarrachea-Magnani, Miguel A. and Str{\'a}nsk{\'y}, Pavel and Lerma-Hern{\'a}ndez, Sergio and Santos, Lea F. and Hirsch, Jorge G.},
  title = {Quantum and Classical Lyapunov Exponents in Atom-Field Interaction Systems},
  journal = {Physical Review Letters},
  volume = {122},
  pages = {024101},
  year = {2019},
  doi = {10.1103/PhysRevLett.122.024101}
}

@article{MaDingYu2020,
  author = {Ma, Yong-Hong and Ding, Quan-Zhen and Yu, Ting},
  title = {Persistent spin squeezing of a dissipative one-axis twisting model embedded in a general thermal environment},
  journal = {Physical Review A},
  volume = {101},
  pages = {022327},
  year = {2020},
  doi = {10.1103/PhysRevA.101.022327}
}

@article{DingZhaoMaChen2021,
  author = {Ding, Quanzhen and Zhao, Peng and Ma, Yonghong and Chen, Yusui},
  title = {Impact of the central frequency of environment on non-Markovian dynamics in piezoelectric optomechanical devices},
  journal = {Scientific Reports},
  volume = {11},
  pages = {1814},
  year = {2021},
  doi = {10.1038/s41598-021-81136-4}
}

@article{ChenDingShiJunYu2020,
  author = {Chen, Yusui and Ding, Quanzhen and Shi, Wufu and Jun, Jing and Yu, Ting},
  title = {Exact entanglement dynamics mediated by leaky optical cavities},
  journal = {Journal of Physics B: Atomic, Molecular and Optical Physics},
  volume = {53},
  pages = {125501},
  year = {2020},
  doi = {10.1088/1361-6455/ab707c}
}

@article{LarkinOvchinnikov1969,
  author = {Larkin, A. I. and Ovchinnikov, Yu. N.},
  title = {Quasiclassical Method in the Theory of Superconductivity},
  journal = {Soviet Physics JETP},
  volume = {28},
  pages = {1200--1205},
  year = {1969}
}

@article{Tuziemski2019,
  author = {Tuziemski, Jan},
  title = {Out-of-time-ordered correlation functions in open systems: A {Feynman--Vernon} influence functional approach},
  journal = {Physical Review A},
  volume = {100},
  pages = {062106},
  year = {2019},
  doi = {10.1103/PhysRevA.100.062106}
}

@article{KirtonKeelingBrandes2019,
  author = {Kirton, Peter and Roses, Michael M. and Keeling, Jonathan and Dalla Torre, Emanuele G.},
  title = {Introduction to the {Dicke} Model: From Equilibrium to Nonequilibrium, and Vice Versa},
  journal = {Advanced Quantum Technologies},
  volume = {2},
  pages = {1800043},
  year = {2019},
  doi = {10.1002/qute.201800043}
}

@article{TiwariBanerjee2023,
  author = {Tiwari, Devvrat and Banerjee, Subhashish},
  title = {Quantum chaos in the {Dicke} model and its variants},
  journal = {Proceedings of the Royal Society A},
  volume = {479},
  pages = {20230431},
  year = {2023},
  doi = {10.1098/rspa.2023.0431}
}

@article{MollabashiRahimiKeshari2025,
  author = {Mollabashi, Ali and Rahimi-Keshari, Saleh},
  title = {Information scrambling in bosonic {Gaussian} dynamics},
  journal = {Physical Review E},
  volume = {112},
  pages = {034213},
  year = {2025},
  doi = {10.1103/wvpl-jdgk}
}

@article{ChenYu2014,
  author = {Chen, Yusui and Yu, Ting},
  title = {Non-{Markovian} dynamics of multiple qubit systems: Exact master equation and quantum trajectories},
  journal = {Physical Review A},
  volume = {90},
  pages = {052104},
  year = {2014},
  doi = {10.1103/PhysRevA.90.052104}
}

@book{GardinerZoller2004,
  author = {Gardiner, Crispin W. and Zoller, Peter},
  title = {Quantum Noise},
  edition = {3},
  publisher = {Springer},
  address = {Berlin},
  year = {2004}
}

@book{WallsMilburn2008,
  author = {Walls, Daniel F. and Milburn, Gerard J.},
  title = {Quantum Optics},
  edition = {2},
  publisher = {Springer},
  address = {Berlin},
  year = {2008}
}

@article{Weedbrook2012,
  author = {Weedbrook, Christian and Pirandola, Stefano and Garc{\'i}a-Patr{\'o}n, Ra{\'u}l and Cerf, Nicolas J. and Ralph, Timothy C. and Shapiro, Jeffrey H. and Lloyd, Seth},
  title = {Gaussian quantum information},
  journal = {Reviews of Modern Physics},
  volume = {84},
  pages = {621--669},
  year = {2012},
  doi = {10.1103/RevModPhys.84.621}
}

@article{Kubo1957,
  author = {Kubo, Ryogo},
  title = {Statistical-Mechanical Theory of Irreversible Processes. I. General Theory and Simple Applications to Magnetic and Conduction Problems},
  journal = {Journal of the Physical Society of Japan},
  volume = {12},
  pages = {570--586},
  year = {1957},
  doi = {10.1143/JPSJ.12.570}
}

@article{GoriniKossakowskiSudarshan1976,
  author = {Gorini, Vittorio and Kossakowski, Andrzej and Sudarshan, E. C. G.},
  title = {Completely positive dynamical semigroups of {N}-level systems},
  journal = {Journal of Mathematical Physics},
  volume = {17},
  pages = {821--825},
  year = {1976},
  doi = {10.1063/1.522979}
}

@article{Lindblad1976,
  author = {Lindblad, Goran},
  title = {On the generators of quantum dynamical semigroups},
  journal = {Communications in Mathematical Physics},
  volume = {48},
  pages = {119--130},
  year = {1976},
  doi = {10.1007/BF01608499}
}

@article{DimerEstienneParkinsCarmichael2007,
  author = {Dimer, F. and Estienne, B. and Parkins, A. S. and Carmichael, H. J.},
  title = {Proposed realization of the {Dicke}-model quantum phase transition in an optical cavity {QED} system},
  journal = {Physical Review A},
  volume = {75},
  pages = {013804},
  year = {2007},
  doi = {10.1103/PhysRevA.75.013804}
}

@article{BaumannGuerlinBrenneckeEsslinger2010,
  author = {Baumann, Kristian and Guerlin, Christine and Brennecke, Ferdinand and Esslinger, Tilman},
  title = {Dicke quantum phase transition with a superfluid gas in an optical cavity},
  journal = {Nature},
  volume = {464},
  pages = {1301--1306},
  year = {2010},
  doi = {10.1038/nature09009}
}

@article{RitschDomokosBrenneckeEsslinger2013,
  author = {Ritsch, Helmut and Domokos, Peter and Brennecke, Ferdinand and Esslinger, Tilman},
  title = {Cold atoms in cavity-generated dynamical optical potentials},
  journal = {Reviews of Modern Physics},
  volume = {85},
  pages = {553--601},
  year = {2013},
  doi = {10.1103/RevModPhys.85.553}
}

@book{BreuerPetruccione2002,
  author = {Breuer, Heinz-Peter and Petruccione, Francesco},
  title = {The Theory of Open Quantum Systems},
  publisher = {Oxford University Press},
  address = {Oxford},
  year = {2002}
}

@article{6fsj-6nc5,
  title = {Quantum speed limit for the out-of-time-ordered correlator from an open-system perspective},
  author = {Tripathy, Devjyoti and Thingna, Juzar and Deffner, Sebastian},
  journal = {Phys. Rev. A},
  volume = {113},
  issue = {1},
  pages = {L010402},
  numpages = {8},
  year = {2026},
  month = {Jan},
  publisher = {American Physical Society},
  doi = {10.1103/6fsj-6nc5},
  url = {https://link.aps.org/doi/10.1103/6fsj-6nc5}
}

@article{98vl-zlcs,
  title = {Analytical determination of multitime correlation functions in quantum chaotic systems},
  author = {Chorbadzhiyska, Yoana R. and Ivanov, Peter A. and Nation, Charlie},
  journal = {Phys. Rev. E},
  volume = {113},
  issue = {4},
  pages = {044208},
  numpages = {21},
  year = {2026},
  month = {Apr},
  publisher = {American Physical Society},
  doi = {10.1103/98vl-zlcs},
  url = {https://link.aps.org/doi/10.1103/98vl-zlcs}
}

@article{PhysRevLett.112.246401,
  title = {Transient Orthogonality Catastrophe in a Time-Dependent Nonequilibrium Environment},
  author = {Schir\'o, Marco and Mitra, Aditi},
  journal = {Phys. Rev. Lett.},
  volume = {112},
  issue = {24},
  pages = {246401},
  numpages = {6},
  year = {2014},
  month = {Jun},
  publisher = {American Physical Society},
  doi = {10.1103/PhysRevLett.112.246401},
  url = {https://link.aps.org/doi/10.1103/PhysRevLett.112.246401}
}

@article{PhysRevB.98.184416,
  title = {Entanglement production and information scrambling in a noisy spin system},
  author = {Knap, Michael},
  journal = {Phys. Rev. B},
  volume = {98},
  issue = {18},
  pages = {184416},
  numpages = {8},
  year = {2018},
  month = {Nov},
  publisher = {American Physical Society},
  doi = {10.1103/PhysRevB.98.184416},
  url = {https://link.aps.org/doi/10.1103/PhysRevB.98.184416}
}

@article{PhysRevB.88.075207,
  title = {Organic magnetoresistance near saturation: Mesoscopic effects in small devices},
  author = {Roundy, R. C. and Vardeny, Z. V. and Raikh, M. E.},
  journal = {Phys. Rev. B},
  volume = {88},
  issue = {7},
  pages = {075207},
  numpages = {5},
  year = {2013},
  month = {Aug},
  publisher = {American Physical Society},
  doi = {10.1103/PhysRevB.88.075207},
  url = {https://link.aps.org/doi/10.1103/PhysRevB.88.075207}
}

@article{Xu:25,
author = {Hui Hui Xu and Feng Ze Cao and E Wu and Yong Hong Ma},
journal = {Opt. Express},
number = {24},
pages = {49996--50006},
publisher = {Optica Publishing Group},
title = {Chaos generation and control in optoelectromechanical systems},
volume = {33},
month = {Dec},
year = {2025},
url = {https://opg.optica.org/oe/abstract.cfm?URI=oe-33-24-49996},
doi = {10.1364/OE.577725},
}

@article{PhysRevD.109.L121902,
  title = {Speed limits to the growth of Krylov complexity in open quantum systems},
  author = {Bhattacharya, Aranya and Nath, Pingal Pratyush and Sahu, Himanshu},
  journal = {Phys. Rev. D},
  volume = {109},
  issue = {12},
  pages = {L121902},
  numpages = {6},
  year = {2024},
  month = {Jun},
  publisher = {American Physical Society},
  doi = {10.1103/PhysRevD.109.L121902},
  url = {https://link.aps.org/doi/10.1103/PhysRevD.109.L121902}
}

@article{PhysRevB.103.174106,
  title = {Strong tunable spin-spin interaction in a weakly coupled nitrogen vacancy spin-cavity electromechanical system},
  author = {Xiong, Wei and Chen, Jiaojiao and Fang, Baolong and Wang, Mingfeng and Ye, Liu and You, J. Q.},
  journal = {Phys. Rev. B},
  volume = {103},
  issue = {17},
  pages = {174106},
  numpages = {8},
  year = {2021},
  month = {May},
  publisher = {American Physical Society},
  doi = {10.1103/PhysRevB.103.174106},
  url = {https://link.aps.org/doi/10.1103/PhysRevB.103.174106}
}

@article{x19r-pzyb,
  title = {Multimode Cavity QED Ising Spin Glass},
  author = {Marsh, Brendan P. and Schuller, David Atri and Ji, Yunpeng and Hunt, Henry S. and Socolof, Giulia Z. and Bowman, Deven P. and Keeling, Jonathan and Lev, Benjamin L.},
  journal = {Phys. Rev. Lett.},
  volume = {135},
  issue = {16},
  pages = {160403},
  numpages = {7},
  year = {2025},
  month = {Oct},
  publisher = {American Physical Society},
  doi = {10.1103/x19r-pzyb},
  url = {https://link.aps.org/doi/10.1103/x19r-pzyb}
}

@article{PhysRevB.100.014415,
  title = {Cavity-mediated dissipative spin-spin coupling},
  author = {Grigoryan, Vahram L. and Xia, Ke},
  journal = {Phys. Rev. B},
  volume = {100},
  issue = {1},
  pages = {014415},
  numpages = {7},
  year = {2019},
  month = {Jul},
  publisher = {American Physical Society},
  doi = {10.1103/PhysRevB.100.014415},
  url = {https://link.aps.org/doi/10.1103/PhysRevB.100.014415}
}

@article{PhysRevResearch.7.L012058,
  title = {Generation and optimization of entanglement between atoms chirally coupled to spin cavities},
  author = {You, Jia-Bin and Kong, Jian Feng and Aghamalyan, Davit and Mok, Wai-Keong and Lim, Kian Hwee and Ye, Jun and Png, Ching Eng and Garc\'{\i}a-Vidal, Francisco J.},
  journal = {Phys. Rev. Res.},
  volume = {7},
  issue = {1},
  pages = {L012058},
  numpages = {6},
  year = {2025},
  month = {Mar},
  publisher = {American Physical Society},
  doi = {10.1103/PhysRevResearch.7.L012058},
  url = {https://link.aps.org/doi/10.1103/PhysRevResearch.7.L012058}
}

@article{PhysRevA.81.062308,
  title = {Single-nuclear-spin cavity QED},
  author = {Takeuchi, Makoto and Takei, Nobuyuki and Doi, Kodai and Zhang, Peng and Ueda, Masahito and Kozuma, Mikio},
  journal = {Phys. Rev. A},
  volume = {81},
  issue = {6},
  pages = {062308},
  numpages = {5},
  year = {2010},
  month = {Jun},
  publisher = {American Physical Society},
  doi = {10.1103/PhysRevA.81.062308},
  url = {https://link.aps.org/doi/10.1103/PhysRevA.81.062308}
}

@article{PhysRevB.104.L100415,
  title = {Tunable cooperativity in coupled spin-cavity systems},
  author = {Liensberger, Lukas and Haslbeck, Franz X. and Bauer, Andreas and Berger, Helmuth and Gross, Rudolf and Huebl, Hans and Pfleiderer, Christian and Weiler, Mathias},
  journal = {Phys. Rev. B},
  volume = {104},
  issue = {10},
  pages = {L100415},
  numpages = {7},
  year = {2021},
  month = {Sep},
  publisher = {American Physical Society},
  doi = {10.1103/PhysRevB.104.L100415},
  url = {https://link.aps.org/doi/10.1103/PhysRevB.104.L100415}
}

@article{PhysRevE.98.012216,
  title = {Bound on the exponential growth rate of out-of-time-ordered correlators},
  author = {Tsuji, Naoto and Shitara, Tomohiro and Ueda, Masahito},
  journal = {Phys. Rev. E},
  volume = {98},
  issue = {1},
  pages = {012216},
  numpages = {7},
  year = {2018},
  month = {Jul},
  publisher = {American Physical Society},
  doi = {10.1103/PhysRevE.98.012216},
  url = {https://link.aps.org/doi/10.1103/PhysRevE.98.012216}
}

@article{PhysRevA.101.053604,
  title = {Classical and quantum chaos in a three-mode bosonic system},
  author = {Rautenberg, Michael and G\"arttner, Martin},
  journal = {Phys. Rev. A},
  volume = {101},
  issue = {5},
  pages = {053604},
  numpages = {12},
  year = {2020},
  month = {May},
  publisher = {American Physical Society},
  doi = {10.1103/PhysRevA.101.053604},
  url = {https://link.aps.org/doi/10.1103/PhysRevA.101.053604}
}

@article{PhysRevB.98.134305,
  title = {Out-of-time-ordered correlators in short-range and long-range hard-core boson models and in the Luttinger-liquid model},
  author = {Lin, Cheng-Ju and Motrunich, Olexei I.},
  journal = {Phys. Rev. B},
  volume = {98},
  issue = {13},
  pages = {134305},
  numpages = {15},
  year = {2018},
  month = {Oct},
  publisher = {American Physical Society},
  doi = {10.1103/PhysRevB.98.134305},
  url = {https://link.aps.org/doi/10.1103/PhysRevB.98.134305}
}

@article{PhysRevB.101.174313,
  title = {Lyapunov growth in quantum spin chains},
  author = {Craps, Ben and De Clerck, Marine and Janssens, Djunes and Luyten, Vincent and Rabideau, Charles},
  journal = {Phys. Rev. B},
  volume = {101},
  issue = {17},
  pages = {174313},
  numpages = {17},
  year = {2020},
  month = {May},
  publisher = {American Physical Society},
  doi = {10.1103/PhysRevB.101.174313},
  url = {https://link.aps.org/doi/10.1103/PhysRevB.101.174313}
}

@article{PhysRevA.81.032314,
  title = {Multilevel Holstein-Primakoff approximation and its application to atomic spin squeezing and ensemble quantum memories},
  author = {Kurucz, Z. and M\o{}lmer, K.},
  journal = {Phys. Rev. A},
  volume = {81},
  issue = {3},
  pages = {032314},
  numpages = {11},
  year = {2010},
  month = {Mar},
  publisher = {American Physical Society},
  doi = {10.1103/PhysRevA.81.032314},
  url = {https://link.aps.org/doi/10.1103/PhysRevA.81.032314}
}

@article{gckn-mc7w,
  title = {Projected Holstein-Primakoff boson representation of quantum spins for spin wave theory},
  author = {Liu, Ke and Xiong, Fangyu and Wang, Fa},
  journal = {Phys. Rev. B},
  volume = {113},
  issue = {2},
  pages = {024413},
  numpages = {12},
  year = {2026},
  month = {Jan},
  publisher = {American Physical Society},
  doi = {10.1103/gckn-mc7w},
  url = {https://link.aps.org/doi/10.1103/gckn-mc7w}
}

@article{PhysRevResearch.2.043243,
  title = {Resummation of the Holstein-Primakoff expansion and differential equation approach to operator square roots},
  author = {Vogl, Michael and Laurell, Pontus and Zhang, Hao and Okamoto, Satoshi and Fiete, Gregory A.},
  journal = {Phys. Rev. Res.},
  volume = {2},
  issue = {4},
  pages = {043243},
  numpages = {11},
  year = {2020},
  month = {Nov},
  publisher = {American Physical Society},
  doi = {10.1103/PhysRevResearch.2.043243},
  url = {https://link.aps.org/doi/10.1103/PhysRevResearch.2.043243}
}

@article{PhysRevA.39.3196,
  title = {Jaynes-Cummings model with intensity-dependent coupling interacting with Holstein-Primakoff SU(1,1) coherent state},
  author = {Bu\ifmmode \check{z}\else \v{z}\fi{}ek, Vladim\'{\i}r},
  journal = {Phys. Rev. A},
  volume = {39},
  issue = {6},
  pages = {3196--3199},
  numpages = {0},
  year = {1989},
  month = {Mar},
  publisher = {American Physical Society},
  doi = {10.1103/PhysRevA.39.3196},
  url = {https://link.aps.org/doi/10.1103/PhysRevA.39.3196}
}

@article{PhysRevA.74.054101,
  title = {Critical property of the geometric phase in the Dicke model},
  author = {Chen, Gang and Li, Juqi and Liang, J.-Q.},
  journal = {Phys. Rev. A},
  volume = {74},
  issue = {5},
  pages = {054101},
  numpages = {3},
  year = {2006},
  month = {Nov},
  publisher = {American Physical Society},
  doi = {10.1103/PhysRevA.74.054101},
  url = {https://link.aps.org/doi/10.1103/PhysRevA.74.054101}
}

@article{PhysRevA.86.012116,
  title = {Dissipative phase transition in a central spin system},
  author = {Kessler, E. M. and Giedke, G. and Imamoglu, A. and Yelin, S. F. and Lukin, M. D. and Cirac, J. I.},
  journal = {Phys. Rev. A},
  volume = {86},
  issue = {1},
  pages = {012116},
  numpages = {21},
  year = {2012},
  month = {Jul},
  publisher = {American Physical Society},
  doi = {10.1103/PhysRevA.86.012116},
  url = {https://link.aps.org/doi/10.1103/PhysRevA.86.012116}
}

@article{PhysRevA.96.032125,
  title = {Effects of counter-rotating-wave terms on the non-Markovianity in quantum open systems},
  author = {Wu, Wei and Liu, Maoxin},
  journal = {Phys. Rev. A},
  volume = {96},
  issue = {3},
  pages = {032125},
  numpages = {10},
  year = {2017},
  month = {Sep},
  publisher = {American Physical Society},
  doi = {10.1103/PhysRevA.96.032125},
  url = {https://link.aps.org/doi/10.1103/PhysRevA.96.032125}
}

@article{PhysRevA.84.043819,
  title = {Importance of counter-rotating coupling in the superfluid-to-Mott-insulator quantum phase transition of light in the Jaynes-Cummings lattice},
  author = {Zheng, Hang and Takada, Yasutami},
  journal = {Phys. Rev. A},
  volume = {84},
  issue = {4},
  pages = {043819},
  numpages = {8},
  year = {2011},
  month = {Oct},
  publisher = {American Physical Society},
  doi = {10.1103/PhysRevA.84.043819},
  url = {https://link.aps.org/doi/10.1103/PhysRevA.84.043819}
}

@article{PhysRevA.96.063820,
  title = {Observing pure effects of counter-rotating terms without ultrastrong coupling: A single photon can simultaneously excite two qubits},
  author = {Wang, Xin and Miranowicz, Adam and Li, Hong-Rong and Nori, Franco},
  journal = {Phys. Rev. A},
  volume = {96},
  issue = {6},
  pages = {063820},
  numpages = {10},
  year = {2017},
  month = {Dec},
  publisher = {American Physical Society},
  doi = {10.1103/PhysRevA.96.063820},
  url = {https://link.aps.org/doi/10.1103/PhysRevA.96.063820}
}

@article{PhysRevResearch.4.023101,
  title = {Lasing and counter-lasing phase transitions in a cavity-QED system},
  author = {Stitely, Kevin C. and Giraldo, Andrus and Krauskopf, Bernd and Parkins, Scott},
  journal = {Phys. Rev. Res.},
  volume = {4},
  issue = {2},
  pages = {023101},
  numpages = {24},
  year = {2022},
  month = {May},
  publisher = {American Physical Society},
  doi = {10.1103/PhysRevResearch.4.023101},
  url = {https://link.aps.org/doi/10.1103/PhysRevResearch.4.023101}
}

@article{PhysRevA.104.013722,
  title = {Beyond the Rabi model: Light interactions with polar atomic systems in a cavity},
  author = {Scala, Giovanni and S\l{}owik, Karolina and Facchi, Paolo and Pascazio, Saverio and Pepe, Francesco V.},
  journal = {Phys. Rev. A},
  volume = {104},
  issue = {1},
  pages = {013722},
  numpages = {11},
  year = {2021},
  month = {Jul},
  publisher = {American Physical Society},
  doi = {10.1103/PhysRevA.104.013722},
  url = {https://link.aps.org/doi/10.1103/PhysRevA.104.013722}
}

@article{PhysRevB.111.104425,
  title = {Second-order correlation and squeezing of photons in cavities with ultrastrong magnon-photon interactions},
  author = {Falch, Vemund and Brataas, Arne and Qaiumzadeh, Alireza},
  journal = {Phys. Rev. B},
  volume = {111},
  issue = {10},
  pages = {104425},
  numpages = {13},
  year = {2025},
  month = {Mar},
  publisher = {American Physical Society},
  doi = {10.1103/PhysRevB.111.104425},
  url = {https://link.aps.org/doi/10.1103/PhysRevB.111.104425}
}

@article{PhysRevB.107.045425,
  title = {Quantum electron transport controlled by cavity vacuum fields},
  author = {Arwas, Geva and Ciuti, Cristiano},
  journal = {Phys. Rev. B},
  volume = {107},
  issue = {4},
  pages = {045425},
  numpages = {9},
  year = {2023},
  month = {Jan},
  publisher = {American Physical Society},
  doi = {10.1103/PhysRevB.107.045425},
  url = {https://link.aps.org/doi/10.1103/PhysRevB.107.045425}
}

@article{PhysRevD.87.024011,
  title = {Retarded Green's function of a Vainshtein system and Galileon waves},
  author = {Chu, Yi-Zen and Trodden, Mark},
  journal = {Phys. Rev. D},
  volume = {87},
  issue = {2},
  pages = {024011},
  numpages = {30},
  year = {2013},
  month = {Jan},
  publisher = {American Physical Society},
  doi = {10.1103/PhysRevD.87.024011},
  url = {https://link.aps.org/doi/10.1103/PhysRevD.87.024011}
}

@article{PhysRevE.70.015601,
  title = {Extracting the Green function from diffuse, equipartitioned waves},
  author = {Malcolm, Alison E. and Scales, John A. and van Tiggelen, Bart A.},
  journal = {Phys. Rev. E},
  volume = {70},
  issue = {1},
  pages = {015601(R)},
  numpages = {4},
  year = {2004},
  month = {Jul},
  publisher = {American Physical Society},
  doi = {10.1103/PhysRevE.70.015601},
  url = {https://link.aps.org/doi/10.1103/PhysRevE.70.015601}
}

@article{PhysRevLett.119.176403,
  title = {Finite Temperature Green's Function Approach for Excited State and Thermodynamic Properties of Cool to Warm Dense Matter},
  author = {Kas, J. J. and Rehr, J. J.},
  journal = {Phys. Rev. Lett.},
  volume = {119},
  issue = {17},
  pages = {176403},
  numpages = {5},
  year = {2017},
  month = {Oct},
  publisher = {American Physical Society},
  doi = {10.1103/PhysRevLett.119.176403},
  url = {https://link.aps.org/doi/10.1103/PhysRevLett.119.176403}
}

@article{PhysRevA.94.062512,
  title = {Coupled-cluster Green's function: Analysis of properties originating in the exponential parametrization of the ground-state wave function},
  author = {Peng, Bo and Kowalski, Karol},
  journal = {Phys. Rev. A},
  volume = {94},
  issue = {6},
  pages = {062512},
  numpages = {10},
  year = {2016},
  month = {Dec},
  publisher = {American Physical Society},
  doi = {10.1103/PhysRevA.94.062512},
  url = {https://link.aps.org/doi/10.1103/PhysRevA.94.062512}
}

@article{PhysRevA.87.052127,
  title = {Non-Markovian fermionic stochastic Schr\"odinger equation for open system dynamics},
  author = {Shi, Wufu and Zhao, Xinyu and Yu, Ting},
  journal = {Phys. Rev. A},
  volume = {87},
  issue = {5},
  pages = {052127},
  numpages = {10},
  year = {2013},
  month = {May},
  publisher = {American Physical Society},
  doi = {10.1103/PhysRevA.87.052127},
  url = {https://link.aps.org/doi/10.1103/PhysRevA.87.052127}
}

\end{document}